\newcounter{myctr}

\documentclass{ws-acs}
\usepackage{url}

\usepackage[utf8]{inputenc}
\usepackage{graphicx}
\usepackage{listings}
\usepackage[english]{babel}
\usepackage{color}
\usepackage{amsmath}
\usepackage{graphicx}
\usepackage{float}
\usepackage{amssymb}
\usepackage{hyperref}
\usepackage{subfig}
\usepackage{booktabs}
\usepackage{url}
\usepackage{listings}
\graphicspath{{images/}}
\graphicspath{{images/},{figures/}}

\ifdefined\anonymous
\usepackage[]{lineno}  \linenumbers 
\fi

\newcommand{\eref}[1]{eq.~(\ref{#1})}

\newcommand{\change}[1]{#1}

\usepackage[draft,inline,nomargin]{fixme} \fxsetup{theme=color} 
\FXRegisterAuthor{cp}{acp}{\color{blue}CP}
\FXRegisterAuthor{jfv}{jafv}{\color{green}JFV}
\FXRegisterAuthor{cg}{cgg}{\color{red}CGG}
\FXRegisterAuthor{jama}{ajama}{\color{green}JAMA}

\begin {document}

\makeatletter
\def\@biblabel#1{[#1]}
\makeatother

\markboth{Morales, Flores, Gershenson, and Pineda}{Statistical properties of rankings in sports and games}

%
\catchline{}{}{}{}{}
%

\title {STATISTICAL PROPERTIES OF RANKINGS IN SPORTS AND GAMES}

\author{\footnotesize JOSÉ ANTONIO MORALES}
\address{Instituto de Física, Universidad Nacional Autónoma de México, Mexico City, 04510, Mexico.\\
jamafcc@ciencias.unam.mx}

\author{\footnotesize JORGE FLORES\footnote{Deceased.}}
\address{Instituto de Física, Universidad Nacional Autónoma de México, Mexico City, 04510, Mexico.}

\author{\footnotesize CARLOS GERSHENSON}
\address{Instituto de Investigaciones en Matemáticas Aplicadas y Sistemas,
Universidad Nacional Autónoma de México, Mexico City, 04510, Mexico \\
Centro de Ciencias de la Complejidad,
Universidad Nacional Autónoma de México, Mexico City, 04510, Mexico \\
Lakeside Labs GmbH, Lakeside Park B04, 9020 Klagenfurt am Wörthersee, Austria\\
cgg@unam.mx}

\author{\footnotesize CARLOS PINEDA}
\address{Instituto de Física, Universidad Nacional Autónoma de México, Mexico City, 04510, Mexico.\\
carlosp@fisica.unam.mx}

\maketitle

\begin{history}
\received{(received date)}
\revised{(revised date)}
\end{history}

\begin{abstract} 
Any collection can be ranked.  Sports and games are common examples of ranked
systems: players and teams are constantly ranked using different methods. The
statistical properties of rankings have been studied for almost a century in a
variety of fields. More recently, data availability has allowed us to study
rank dynamics: how elements of a ranking change in time. Here, we study the
rank distributions and rank dynamics of twelve datasets from different sports
and games. To study rank dynamics, we consider measures we have defined
previously: rank diversity, change probability, rank entropy, and rank
complexity. We also introduce a new measure that we call ``system closure''
that reflects how many elements enter or leave the rankings in time. We  use a
random walk model to reproduce the observed rank dynamics, showing that a
simple mechanism can generate similar statistical properties as the ones
observed in the datasets. Our results show that, while rank distributions vary
considerably for different rankings, rank dynamics have similar behaviors,
independently \change{of} the nature and competitiveness of the sport or game and its
ranking method. Our results also suggest that our measures of rank dynamics are
general and applicable for complex systems of different natures. 

\end{abstract} 


\section{Introduction} 
\noindent

Ranking in sports has been a common instrument to classify players and teams.
The best team or player is given rank $k=1$, the second one has $k=2$, and so
on.  In each sport, different rules are used to rank, as we show here with two
examples. We first mention the FIFA ranking of national football teams. This
uses the Elo method~\cite{fifa}, that adds or subtracts points for individual
matches from a team's existing point total. If $S_b$ denotes the points a
team has before a given match, the new number of points $S$ is obtained adding
to $S_b$ the quantity $I(W-W_e)$. The value of $I$ varies between 5 and 60
according to the importance of the match, $W$ is the result of the match (being
equal to 1 if the team wins, 0.5 for a draw and 0 if the team is defeated), and
$W_e$ refers to the expected result of the match; the FIFA classification uses
the formula $W_e=1/(10^{-d_r/600}+1)$ with $d_r$ equal to the difference in
$S_b$ of the two teams. As a second example, we refer to the ranking used by
the Global Poker Index (GPI)~\cite{globalPokerInd}. In this case, only
tournaments open to the public are taken into account if they fulfill the
following characteristics: at least 32 players participate, bets must be larger
than 1 dollar, and the tournament should have taken place within the 36 months
before the ranking was made. The score obtained by each player is calculated
according to the position obtained in the tournaments and the amount of money
earned. Older tournaments have a higher weight. These two examples illustrate
how rankings can be constructed. Beyond sports, an arbitrary variable can be
used to order a set of elements into a ranking.

Less attention has been given to the way rank evolves in time. Still, with the
recent data availability, studies of rank dynamics began to
appear~\cite{batty2006rank,blumm2012dynamics,PhysRevX.3.021006,doi:10.1142/S021952591750014X}.  A few years
ago, the first complex system we investigated \cite{cocho2015rank} was related
to languages and the time evolution of word usage was characterized by the rank
diversity, which is equal to the number of different words with rank $k$ in a
time interval $\Delta t$. This analysis was extended in~\cite{Ngramas} to the
case of $N$-grams, that is, groups of $N$ words. In order to understand the
hierarchical formation of word usage, George Zipf~\cite{zipf1932selected,ZipfRnd2002}
proposed that the frequency $f$ of a word is related to the word rank $k$ as
$f\sim 1/k$.  This is known as Zipf's law and is an example of a rank
distribution.  In~\cite{morales2016generic} several complex
systems sports and games were considered. The
disciplines we took into account were football (national teams and clubs), golf
and tennis (for male competitors) as well as poker and chess. In all cases we
considered the distribution of scores versus ranks, that is, the rank
distribution version for these systems, a generalization of the Zipf's law to
the context of sports and games. The rank diversity, which is equal to the
number of different players or teams with rank $k$ in a time interval $\Delta
t$, was also calculated. The random walker model introduced
previously~\cite{cocho2015rank} was used to obtain the observed rank
properties. The rank distributions differs according to the sport, game or
language discussed, but in all cases the rank diversity $d(k)$ was found to be
a log-normal distribution.

In this paper, the analysis of sports and games is extended, considering now
twelve different cases, see
Table~\ref{tab:ranking_data}. Besides the rank diversity, we also obtain for
games and sports the rank change probability, the rank entropy and the rank
complexity previously studied for $N$-grams~\cite{Ngramas}. We also use the
random walker model to fit the rank diversity data.

\begin{tiny} \begin{table*}[h!] 
\footnotesize
\resizebox{\textwidth}{!}{ 
\begin{tabular}{ p{4.0cm} p{3.9cm} p{3.0cm} p{1.2cm} c }
\toprule
Sport/game & Data Source & Time Period & Timescale & \#players/teams \\
\toprule
 \noalign{\smallskip}
 Chess players* \newline (female) & F\'ed\'eration Internationale des \'Echecs (FIDE-F)~\cite{worldChessFed} & Jul 2012 -- Apr 2016 & Monthly & 12681\\
 \midrule
 \noalign{\smallskip}
 Chess players \newline (male) & F\'ed\'eration Internationale des \'Echecs (FIDE-M)~\cite{worldChessFed} & Jul 2012 -- Apr 2016 & Monthly & 13500\\
 \midrule
 \noalign{\smallskip}
 Football teams & Football Club World \newline Ranking (FCWR-C)~\cite{footballClubWorld} & Feb 1st, 2012 -- \newline Dec 29th, 2014 & Weekly & 850\\
 \midrule
 \noalign{\smallskip}
  National football teams & F\'ed\'eration Internationale de Football Association \newline (FIFA)~\cite{fifa} & Jul 2010 -- Dec 2015 & Monthly & 150\\
  \midrule
  \noalign{\smallskip}
  Football Scorers in Clubs*  & Football Club World \newline Ranking (FCWR-G)~\cite{footballClubWorld} & Week 33, 2016 -- \newline Week 33 2017 & Weekly & 400 \\
  \midrule
  \noalign{\smallskip}
  Golf players & Official World Golf Ranking (OWGR)~\cite{officialWorldGolf} & Sept 10th, 2000 -- \newline Apr 19th, 2015 & Weekly & 1000\\
  \midrule
   \noalign{\smallskip}
 Racers in the \newline Busch Grand Nation \newline Tournament* & National Association for Stock Car Auto Racing (NASCAR-B)~\cite{nascar} &  1982 -- 2015 & Annual & 76 \\
 \midrule
 \noalign{\smallskip}
 Racers in the \newline Winston Cup Grand \newline National Tournament* & National Association for Stock Car Auto Racing (NASCAR-W)~\cite{nascar} & 1979 -- 2013 & Annual & 50 \\
 \midrule
  \noalign{\smallskip}
  Poker players & Global Poker Index \newline (GPI)~\cite{globalPokerInd} & Jul 25th, 2012 -- \newline Jun 10th, 2015 & Weekly & 1799\\
  \midrule
  \noalign{\smallskip}
  Snowboard riders* & World Snowboarding \newline (WSD)~\cite{snowBoarding} & January 5th, 2015 -- March 26th, 2018 & Weekly & 1413 \\
  \midrule
 \noalign{\smallskip}
 Tennis players \newline (male) & Association of Tennis \newline Professionals (ATP)~\cite{atpWorldTour} & May 5th, 2003 -- \newline Dec 27th, 2010 & Weekly & 1600\\
 \midrule
\noalign{\smallskip}
 Videogame earnings* & E-Sports Earnings \newline(ESE)~\cite{videogames} & 2003 -- 2016 & Annual  & 400 \\
\bottomrule

\end{tabular}
}
\caption{ {\bf Ranking data for the twelve datasets studied in this paper.} }
\label{tab:ranking_data}

\end{table*}
\end{tiny} 

\section{Rank Distribution} 

\noindent
Considering birth and death processes, we obtained in reference~\cite{cocho2015rank}
four generalizations of Zipf's law: 
\begin{equation}
m_1(k)=\mathcal{N} \frac{1}{k^a}, \quad\quad m_2(k)=\mathcal{N} \frac{\exp(-bk)}{k^a},
  \label{eqn:1}
\end{equation}
\begin{equation}
m_3(k)=\mathcal{N} \frac{(N+1-k)^q}{k^a}, \quad\quad m_4(k)=f(k)=\mathcal{N}\frac{(N+1-k)^q\exp(-bk)}{k^a};
  \label{eqn:3}
\end{equation}
a fifth model given by the double Zipf law~\cite{PhysRevX.3.021006},
\begin{equation}
m_5(k)=\mathcal{N} \left\{ \begin{array}{lcc}
             \frac{1}{k^a}, &  k\leq k_c \\
             \\ \frac{k_c^{a'-a}}{k^{a'}} & k>k_c,
             \end{array}
\right.
\label{eqn:5}
\end{equation}
\noindent
was also taken into account. 

Here $N$ is the number of words considered, $\mathcal{N}$ is a normalization
constant, $a$, $b$, $q$ and $k_c$ are parameters to be fixed fitting the data.
$m_1$ is Zipf's law, a power law distribution. If plotted in $\log \log$ scale,
it is a straight line with steepness $-a$.  $m_2$ and $m_3$ are called in the
literature the gamma and beta distributions,
respectively, while $m_4$ is a combination of both. In the linguistic case, Gerlach and Altman
\cite{PhysRevX.3.021006} introduced model $m_5$ which is the concatenation of
two power laws, one corresponding to words which have high frequency and the
other to those with low frequency. To find out which distribution fits the data
better, we shall use two coefficients: the coefficient of determination $R^2$
\cite{li2010fitting} and the Kolmogorov-Smirnov index $p$
\cite{kolmogorov1933sulla,clauset2009power}, which is only applicable when
dealing with probability distributions, this being our case. The closer $R^2$
is to unity, the model fits the data better. On the other hand, when $p< 0.1$
the model can be discarded. 

\begin{figure} 
\centering
\includegraphics{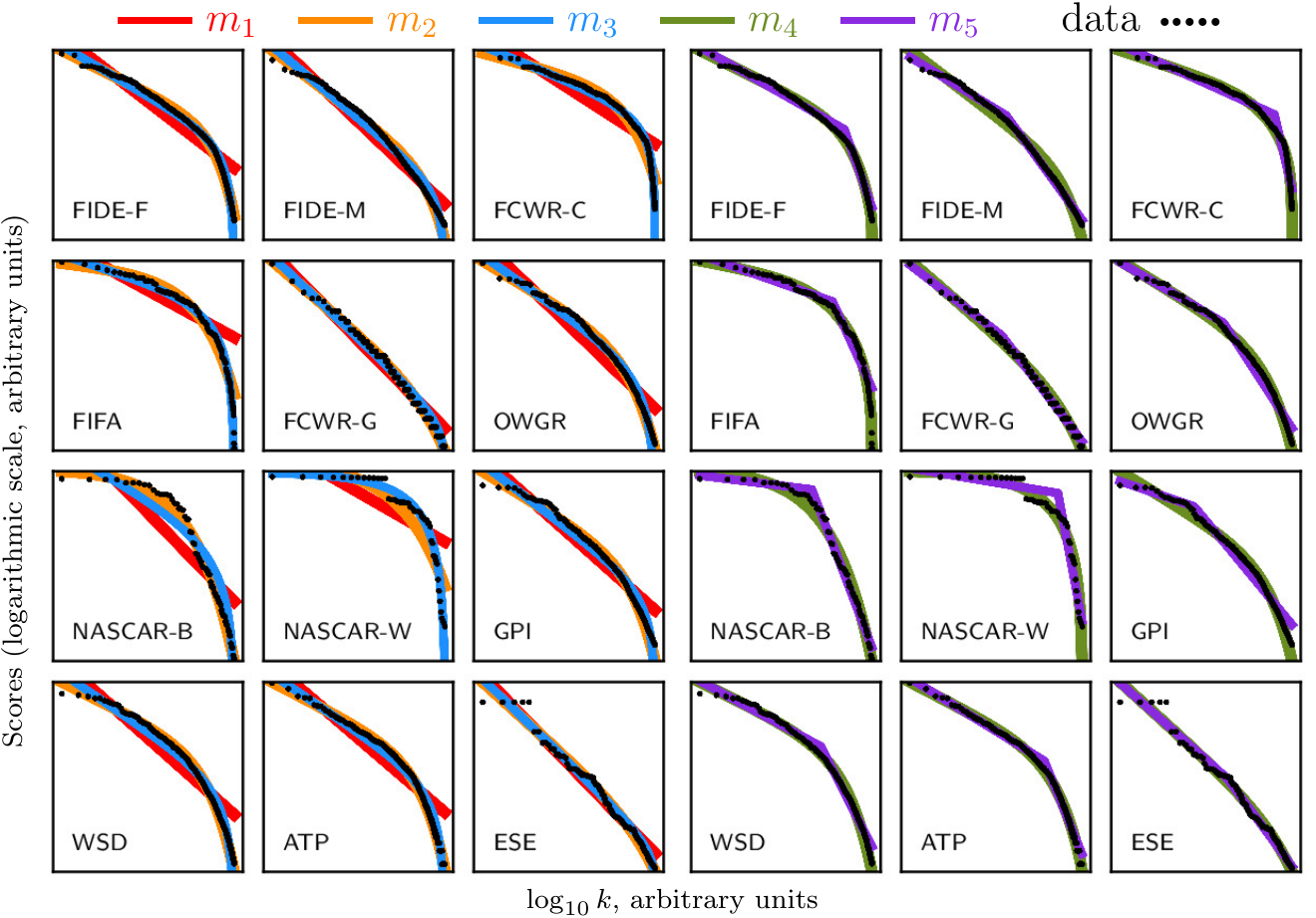}
\caption{\textbf{Comparison of ranking data for the twelve rankings
with models $m_{1,2,3,4,5}$.} Zipf's law ($m_1$) does not reproduce
satisfactorily the data in any case, while the gamma distribution ($m_2$) is
appropriate in some of them. Beta distribution ($m_3$) falls abruptly for large
$k$ while distribution $m_4$ fits the data reasonably well.
In particular, the cases NASCAR-B, NASCAR-W and ESE show an abrupt change that can be well explained by $m_5$. The range of the axes is such that all data points and fits are 
appropriately displayed. 
}
\label{fig:ranktest}
\end{figure}

The comparisons of the five models with the observed data are shown for the
twelve sports and games in Fig.~\ref{fig:ranktest}. The data correspond to the following time slices: FIDE-F (April
2016), FIDE-M (April 2016), FCWR-C (week 53 of 2014), FIFA (June 2017),
FCWR-G (week 33 of 2017), OWGR(May 21st of 2017), NASCAR-B (2015),
NASCAR-W (2013), GPI (May 31st of 2017), WSD (April 26, 2018), ATP (December 27,
2010), and ESE (2016). The values of the parameters of the different distributions
$m_i$ that fit the data better are given in Table \ref{tab:fitting_rank_data}. 

\begin{table} 
\centering
\resizebox{\textwidth}{!}{ 
\begin{tabular}{|l|c| c | c | c |c | c | c |c|c|c|c|c|c|c|c|}\hline 
& \multicolumn{2}{|c|}{\textbf{Model $m_1$}}  & \multicolumn{3}{|c|}{\textbf{Model $m_2$}}  & \multicolumn{3}{|c|}{\textbf{Model $m_3$}}\\ \hline 
&  $\log \mathcal{N}$ & $a$ & $\log \mathcal{N}$ & $a$ & $b$ & $\log \mathcal{N}$ & $a$ & $q$ \\ \hline 
\textbf{FIDE-F} & $ 3.452 $ & $ 4.33\times 10^{-2}$  & $ 3.432 $ & $ 2.78\times 10^{-2}$  & $ 1.94\times 10^{-5}$  & $ 3.021 $ & $ 3.3\times 10^{-2}$  & $ 0.102 $ \\ \hline 
\textbf{FIDE-M} & $ 3.468 $ & $ 2.49\times 10^{-2}$  & $ 3.461 $ & $ 1.99\times 10^{-2}$  & $ 6.27\times 10^{-6}$  & $ 3.327 $ & $ 2.19\times 10^{-2}$  & $ 3.3\times 10^{-2}$  \\ \hline 
\textbf{FCWR-C} & $ 4.522 $ & $ 0.529 $ & $ 4.242 $ & $ 0.219 $ & $ 3.06\times 10^{-3}$  & $ 2.191 $ & $ 0.341 $ & $ 0.733 $ \\ \hline 
\textbf{FIFA} & $ 3.418 $ & $ 0.407 $ & $ 3.229 $ & $ 8.37\times 10^{-2}$  & $ 1.29\times 10^{-2}$  & $ 1.29 $ & $ 0.235 $ & $ 0.875 $ \\ \hline 
\textbf{FCWR-G} & $ 1.764 $ & $ 0.239 $ & $ 1.736 $ & $ 0.2 $ & $ 8.77\times 10^{-4}$  & $ 1.527 $ & $ 0.221 $ & $ 8.54\times 10^{-2}$  \\ \hline 
\textbf{OWGR} & $ 1.388 $ & $ 0.691 $ & $ 1.145 $ & $ 0.429 $ & $ 2.19\times 10^{-3}$  & $ -1.958 $ & $ 0.511 $ & $ 1.035 $ \\ \hline 
\textbf{NASCAR-B} & $ 3.673 $ & $ 1.031 $ & $ 3.185 $ & $ 6.1\times 10^{-10}$  & $ 6.37\times 10^{-2}$  & $ 1.577 $ & $ 0.56 $ & $ 0.965 $ \\ \hline 
\textbf{NASCAR-W} & $ 3.943 $ & $ 0.964 $ & $ 3.627 $ & $ 1.58\times 10^{-10}$  & $ 9.48\times 10^{-2}$  & $ 0.137 $ & $ 5.54\times 10^{-8}$  & $ 1.943 $ \\ \hline 
\textbf{GPI} & $ 3.652 $ & $ 0.178 $ & $ 3.598 $ & $ 0.122 $ & $ 4.13\times 10^{-4}$  & $ 2.902 $ & $ 0.152 $ & $ 0.222 $ \\ \hline 
\textbf{WSD} & $ 3.323 $ & $ 0.554 $ & $ 3.094 $ & $ 0.31 $ & $ 1.94\times 10^{-3}$  & $ 0.228 $ & $ 0.423 $ & $ 0.942 $ \\ \hline 
\textbf{ATP} & $ 4.511 $ & $ 1.042 $ & $ 4.117 $ & $ 0.626 $ & $ 3.18\times 10^{-3}$  & $ -1.467 $ & $ 0.817 $ & $ 1.795 $ \\ \hline 
\textbf{ESE} & $ 6.551 $ & $ 0.679 $ & $ 6.485 $ & $ 0.589 $ & $ 2.02\times 10^{-3}$  & $ 6.001 $ & $ 0.636 $ & $ 0.198 $ \\ \hline 
\end{tabular} 
}
\centering
\resizebox{\textwidth}{!}{ 
\begin{tabular}{|l|c| c | c | c |c | c | c |c|c|c|c|c|c|c|c|c|c|}\hline 
 &  \multicolumn{4}{|c|}{\textbf{Model $m_4$}}  & \multicolumn{4}{|c|}{\textbf{Model $m_5$}} \\ \hline 
&  $ \log \mathcal{N}$ & $a$ & $b$ & $q$ & $\log \mathcal{N}$ & $a$ &$a'$ & $\log k_c$ \\ \hline 
\textbf{FIDE-F} & $ 3.392 $ & $ 2.81\times 10^{-2}$  & $ 9.83\times 10^{-3}$  & $ 1.79\times 10^{-5}$  & $ 3.436 $ & $ 3.19\times 10^{-2}$  & $ 0.158 $ & $ 2.8\times 10^{3}$  \\ \hline 
\textbf{FIDE-M} & $ 3.461 $ & $ 1.99\times 10^{-2}$  & $ 6.66\times 10^{-13}$  & $ 6.27\times 10^{-6}$  & $ 3.457 $ & $ 1.58\times 10^{-2}$  & $ 3.59\times 10^{-2}$  & $ 2.03\times 10^{2}$  \\ \hline 
\textbf{FCWR-C} & $ 2.937 $ & $ 0.27 $ & $ 0.458 $ & $ 1.4\times 10^{-3}$  & $ 4.357 $ & $ 0.371 $ & $ 3.472 $ & $ 4.27\times 10^{2}$  \\ \hline 
\textbf{FIFA} & $ 2.338 $ & $ 0.132 $ & $ 0.397 $ & $ 7.83\times 10^{-3}$  & $ 3.308 $ & $ 0.263 $ & $ 1.64 $ & $ 59.585 $ \\ \hline 
\textbf{FCWR-G} & $ 1.736 $ & $ 0.2 $ & $ 3.88\times 10^{-10}$  & $ 8.77\times 10^{-4}$  & $ 1.721 $ & $ 0.18 $ & $ 0.309 $ & $ 24.494 $ \\ \hline 
\textbf{OWGR} & $ 1.145 $ & $ 0.429 $ & $ 1.19\times 10^{-9}$  & $ 2.19\times 10^{-3}$  & $ 1.125 $ & $ 0.416 $ & $ 1.094 $ & $ 69.565 $ \\ \hline 
\textbf{NASCAR-B} & $ 3.185 $ & $ 3.5\times 10^{-11}$  & $ 5.86\times 10^{-8}$  & $ 6.37\times 10^{-2}$  & $ 3.112 $ & $ 0.125 $ & $ 2.626 $ & $ 16.795 $ \\ \hline 
\textbf{NASCAR-W} & $ 0.685 $ & $ 2.06\times 10^{-9}$  & $ 1.644 $ & $ 1.65\times 10^{-2}$  & $ 3.524 $ & $ 0.284 $ & $ 8.886 $ & $ 28.962 $ \\ \hline 
\textbf{GPI} & $ 3.598 $ & $ 0.122 $ & $ 6.68\times 10^{-12}$  & $ 4.13\times 10^{-4}$  & $ 3.561 $ & $ 6.8\times 10^{-2}$  & $ 0.242 $ & $ 25.385 $ \\ \hline 
\textbf{WSD} & $ 3.094 $ & $ 0.31 $ & $ 2.29\times 10^{-8}$  & $ 1.94\times 10^{-3}$  & $ 3.104 $ & $ 0.338 $ & $ 1.221 $ & $ 1.64\times 10^{2}$  \\ \hline 
\textbf{ATP} & $ 4.018 $ & $ 0.628 $ & $ 3.12\times 10^{-2}$  & $ 3.14\times 10^{-3}$  & $ 4.2 $ & $ 0.747 $ & $ 3.004 $ & $ 3.19\times 10^{2}$  \\ \hline 
\textbf{ESE} & $ 6.485 $ & $ 0.589 $ & $ 3.7\times 10^{-9}$  & $ 2.02\times 10^{-3}$  & $ 6.474 $ & $ 0.583 $ & $ 0.89 $ & $ 42.08 $ \\ \hline 
\end{tabular} 
}
\caption{Values of the parameters in $m_i$ that fit the data better for all the datasets considered here.} 
\label{tab:fitting_rank_data} 
\end{table}  

The average values, taken over all time slices available of $R^2$ and the
standard deviation $\sigma_{R^2}$ are given in Table~\ref{table:goodnes:of:fits1}. None of the distributions fit the data for all
sports and games; in particular, Zipf's law ($m_1$) is never the best.
Furthermore, using the values of the time average of the Kolmogorov-Smirnov
index $p$, also given in Table~\ref{table:goodnes:of:fits1}, one can conclude that
Zipf's law can be discarded. From the data for FIDE-F, FIDE-M, FCWR-C and GPI,
one can see that none of the models describes the way scores are distributed.
We should also mention that when models $m_2$ or $m_3$ work, so does model
$m_4$. If model $m_3$ is acceptable, as for FIFA and NASCAR-W, one can conclude
that for large values of $k$ the distribution falls abruptly, indicating that
the less able competitors receive a very small amount of points from these
federations compared to those given to the competitors who had the best
performance. When the double Zipf model $m_5$ is adequate, as for FCWR-G,
NASCAR-B, NASCAR-W, ATP and ESE, there exist two independent regimes
characterized by the exponent of the power laws. The first regime corresponds
to the better competitors and the second one to those with poor performance. As
mentioned in reference~\cite{PhysRevX.3.021006}, those belonging to the second
regime do not affect the rank of the competitors in the first regime.

\change{In Table~\ref{table:goodnes:of:fits1}, some models have $p$ index values below the threshold 0.1 while still having a high $R^2$. This is because $p$ quantifies whether a dataset has a distribution similar to a model $m_i$, while $R^2$ only measures how good a dataset fits a function defined by $m_i$. Thus, if most data points fall close to $m_i$, $R^2$ will be high. However, if the \emph{distribution} of those points is not close to the one expected by $m_i$, then $p<0.1$.}


\begin{table}[htbp]
\centering

\resizebox{\textwidth}{!}{ 
\begin{tabular}{l c c c c c c} 
\toprule 
     & & $m_1$ & $m_2$ & $m_3$ & $m_4$ & $m_5$ \\ 
\midrule 
\midrule 
\textbf{FIDE-F} &  $\langle R^2 \rangle$ & 0.445$\pm$0.047 & 0.981$\pm$0.016 & 0.89$\pm$0.031 & 0.995$\pm$0.002 & 0.965$\pm$0.011\\ 
    &  $\langle p \rangle$ & 0.0$\pm$0.0 & 0.0$\pm$0.0 & 0.0$\pm$0.0 & 0.0$\pm$0.0 & 0.036$\pm$0.004\\ 
\hline 
\textbf{FIDE-M} &  $\langle R^2 \rangle$ & 0.778$\pm$0.007 & 0.936$\pm$0.005 & 0.657$\pm$0.017 & 0.936$\pm$0.005 & 0.991$\pm$0.001\\ 
    &  $\langle p \rangle$ & 0.0$\pm$0.0 & 0.0$\pm$0.0 & 0.0$\pm$0.0 & 0.0$\pm$0.0 & 0.001$\pm$0.001\\ 
\hline 
\textbf{FCWR-C} &  $\langle R^2 \rangle$ & 0.727$\pm$0.028 & 0.987$\pm$0.005 & 0.982$\pm$0.005 & 0.997$\pm$0.001 & 0.948$\pm$0.01\\ 
    &  $\langle p \rangle$ & 0.002$\pm$0.002 & 0.0$\pm$0.0 & 0.014$\pm$0.017 & 0.002$\pm$0.006 & 0.001$\pm$0.001\\ 
\hline 
\textbf{FIFA} &  $\langle R^2 \rangle$ & 0.763$\pm$0.021 & 0.987$\pm$0.004 & 0.993$\pm$0.004 & 0.997$\pm$0.001 & 0.97$\pm$0.005\\ 
    &  $\langle p \rangle$ & 0.0$\pm$0.0 & 0.011$\pm$0.012 & 0.415$\pm$0.32 & 0.488$\pm$0.351 & 0.074$\pm$0.038\\ 
\hline 
\textbf{FCWR-G} &  $\langle R^2 \rangle$ & 0.973$\pm$0.013 & 0.992$\pm$0.003 & 0.985$\pm$0.006 & 0.992$\pm$0.003 & 0.992$\pm$0.002\\ 
    &  $\langle p \rangle$ & 0.047$\pm$0.071 & 0.759$\pm$0.234 & 0.032$\pm$0.052 & 0.741$\pm$0.229 & 0.317$\pm$0.206\\ 
\hline 
\textbf{OWGR} &  $\langle R^2 \rangle$ & 0.65$\pm$0.132 & 0.985$\pm$0.014 & 0.971$\pm$0.011 & 0.985$\pm$0.012 & 0.972$\pm$0.009\\ 
    &  $\langle p \rangle$ & 0.0$\pm$0.0 & 0.612$\pm$0.242 & 0.0$\pm$0.0 & 0.527$\pm$0.337 & 0.093$\pm$0.157\\ 
\hline 
\textbf{NASCAR-B} &  $\langle R^2 \rangle$ & 0.392$\pm$0.218 & 0.971$\pm$0.017 & 0.853$\pm$0.068 & 0.974$\pm$0.015 & 0.976$\pm$0.013\\ 
    &  $\langle p \rangle$ & 0.001$\pm$0.002 & 0.332$\pm$0.293 & 0.008$\pm$0.025 & 0.299$\pm$0.293 & 0.258$\pm$0.226\\ 
\hline 
\textbf{NASCAR-W} &  $\langle R^2 \rangle$ & 0.469$\pm$0.162 & 0.845$\pm$0.068 & 0.952$\pm$0.022 & 0.95$\pm$0.025 & 0.932$\pm$0.066\\ 
    &  $\langle p \rangle$ & 0.021$\pm$0.024 & 0.005$\pm$0.018 & 0.153$\pm$0.24 & 0.124$\pm$0.192 & 0.189$\pm$0.16\\ 
\hline 
\textbf{GPI} &  $\langle R^2 \rangle$ & 0.796$\pm$0.014 & 0.977$\pm$0.005 & 0.938$\pm$0.005 & 0.977$\pm$0.005 & 0.982$\pm$0.007\\ 
    &  $\langle p \rangle$ & 0.0$\pm$0.0 & 0.0$\pm$0.0 & 0.0$\pm$0.0 & 0.0$\pm$0.0 & 0.0$\pm$0.001\\ 
\hline 
\textbf{WSD} &  $\langle R^2 \rangle$ & 0.799$\pm$0.016 & 0.989$\pm$0.003 & 0.954$\pm$0.006 & 0.989$\pm$0.003 & 0.972$\pm$0.004\\ 
    &  $\langle p \rangle$ & 0.0$\pm$0.0 & 0.115$\pm$0.148 & 0.0$\pm$0.0 & 0.118$\pm$0.154 & 0.0$\pm$0.0\\ 
\hline 
\textbf{ATP} &  $\langle R^2 \rangle$ & 0.272$\pm$0.211 & 0.982$\pm$0.013 & 0.88$\pm$0.067 & 0.982$\pm$0.013 & 0.965$\pm$0.029\\ 
    &  $\langle p \rangle$ & 0.001$\pm$0.003 & 0.433$\pm$0.232 & 0.0$\pm$0.0 & 0.593$\pm$0.411 & 0.106$\pm$0.11\\ 
\hline 
\textbf{ESE} &  $\langle R^2 \rangle$ & 0.826$\pm$0.104 & 0.94$\pm$0.043 & 0.903$\pm$0.06 & 0.919$\pm$0.105 & 0.965$\pm$0.037\\ 
    &  $\langle p \rangle$ & 0.018$\pm$0.026 & 0.074$\pm$0.11 & 0.023$\pm$0.056 & 0.064$\pm$0.112 & 0.168$\pm$0.243\\ 
\hline 
\end{tabular} 
}
\caption{Averages and standard deviations for the goodness of fit $R^2$ and \change{Kolmogorov-Smirnov index} $p$.}  
\label{table:goodnes:of:fits1} 
\end{table}


\section{Rank Dynamics} 
\begin{figure}[h!] 
\centering
\includegraphics{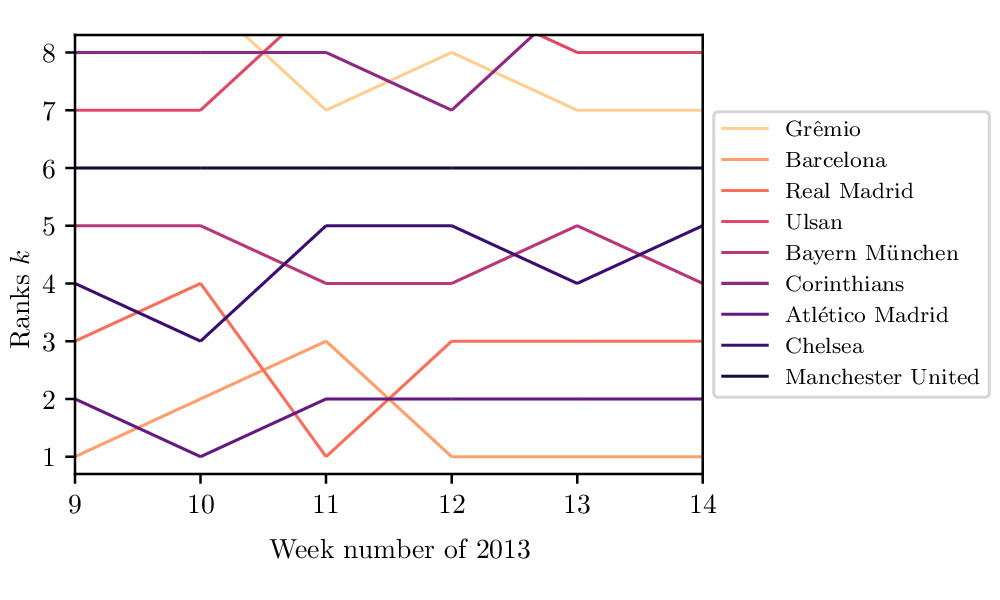}
\caption{\textbf{Rank time evolution of the nine teams that at a certain week
were the top 8.} Example of ranking dynamics. We consider football clubs world
wide (FCWR-C) from week 9th to week 14th of 2013. 
}
\label{fig:example}
\end{figure}

\begin{figure} 
\centering
\includegraphics[width=0.6\textwidth]{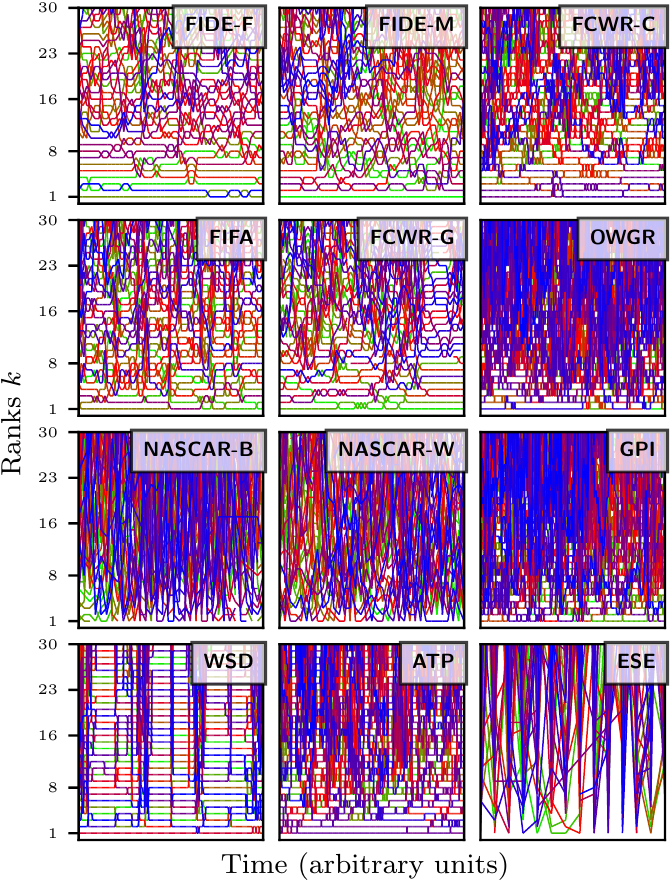}
\caption{Players and teams of different sports and games change their ranks over time. Only best 30 shown.}
\label{fig:spaguettis}
\end{figure}

We shall now study the time evolution of the rank occupancy. In Fig.~\ref{fig:example} we plot $k$ as a function of time for the top 8 football
clubs. These so-called 'spaghetti' curves show how elements --- individuals or
teams --- change their rank in time. 
Fig.~\ref{fig:spaguettis} shows more detailed spaghetti curves. In several, the
following behavior is observed: low ranks change less than higher ranks. In
other words, change of the best is slower than that of the rest. However, this
is less clear for other sports. Since a purely visual inspection might be
misleading, a more formal measure is desirable.

\begin{figure}[h!] 
\centering
\includegraphics[width=\textwidth]{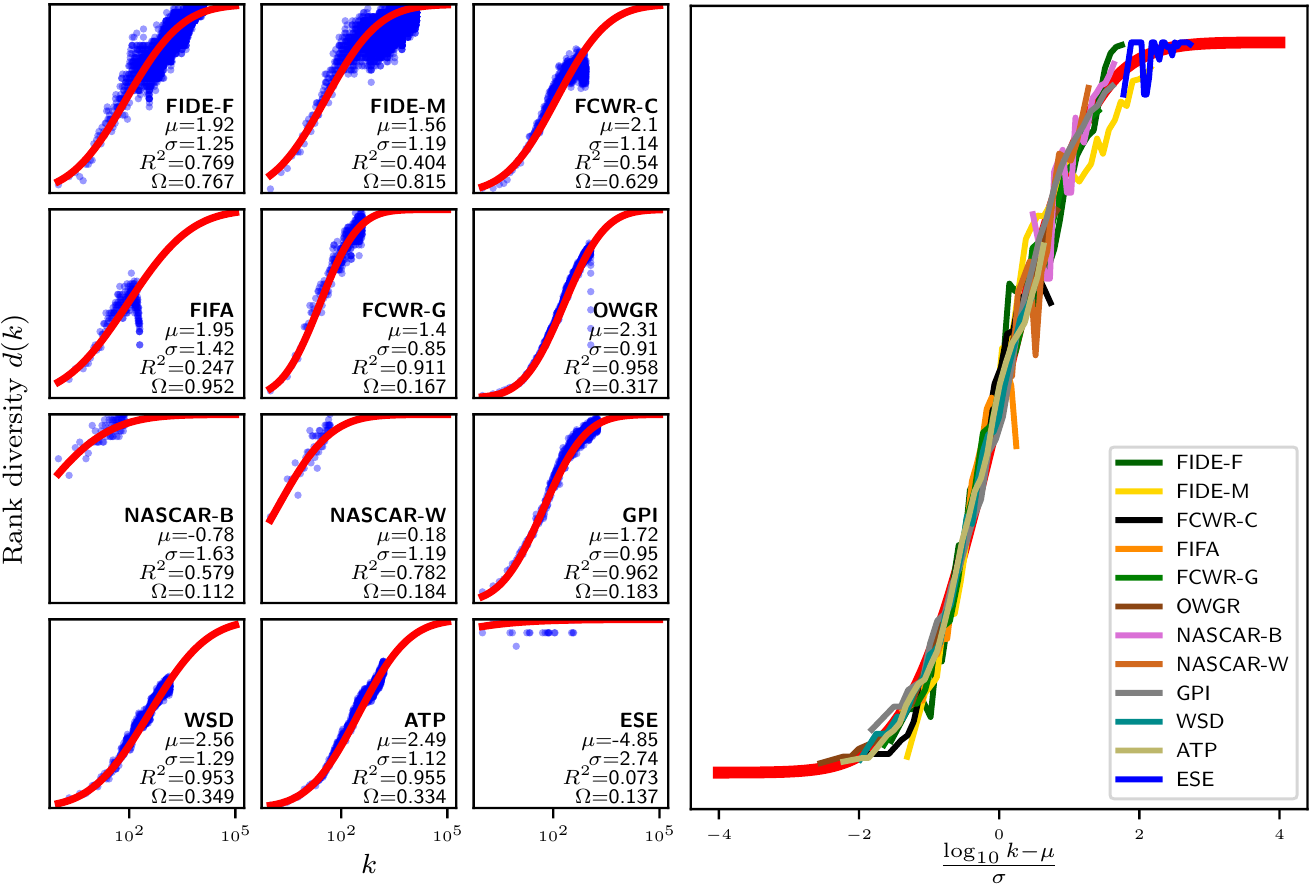}
\caption{(left) Rank diversity plots (blue dots) for the twelve datasets considered
here, together with a fitted sigmoid $\Phi$ (equation~\ref{eqn:Phi}) (red curve). The mean $\mu$, standard deviation $\sigma$, and squared error $R^2$ of $\Phi$ are also shown, together with the closure $\Omega$.
(right) If the variable $(\log_{10}{k}-\mu)/\sigma$ is used,  the rank diversity $d(k)$
coincides well for all datasets. The values of $d(k)$ are always in the interval $[0,1]$.
 }
\label{fig:normalized}
\end{figure}

For this purpose, we can calculate the rank diversity introduced recently 
\cite{cocho2015rank}. If for each sport or game we
have ranking information for $T$ times $t_1,t_2,...,t_T$, we define $X(k,t_i)$
as the element (player or team) with rank $k$ at time $t_i$. If we denote by
$X(k)=\{X(k,t_1),X(k,t_2),...,X(k,t_T)   \}$ the set of elements that have rank
$k$ at all times, the rank diversity is defined as:
\begin{equation}
d(k)=\frac{|X(k)|}{T}
\end{equation}
where $|X(k)|$ is the cardinality of the set $X(k)$. \change{Notice that
diversity will range from $1/T$ (when for a given rank, 
elements are different for each time step) to 1 (when only one element
occupies a given rank).} \change{For example, for the data presented in 
figure \ref{fig:example}, $T=6$, $d(1)=1/2$, and $d(6)=1/6$.} The rank diversity for
several sports and games is presented in figure \ref{fig:normalized} (left).

We will now introduce what we shall call the \emph{system closure}, defined below. If
the elements of the system are always the same we will say that the system is
closed. This happens, for example, for the classification of football national
teams: few countries enter and leave the rankings, as most teams are always considered. To quantify closure we will introduce the closure index $\Omega$. Let us
consider an arbitrary integer $N$ and take into account only those elements
with ranking $k$ between $1$ and $N.$ We denote by $\Gamma$ the number of
elements that at time $t$ had a rank smaller than $N$; therefore, $\Gamma \geq N$. \change{In other words, $N$ is the size of the ranking, \emph{e.g.}, for the ``top 100'', $N=100$, while $\Gamma$ indicates how many elements have passed through the ranking during time $t$.}
The closure index is then 
\begin{equation}
0<\Omega=\frac{N}{\Gamma} \leq 1.
\end{equation}
When all elements are the same for all times, the system is completely closed
and $\Gamma=N \Rightarrow \Omega =1$\change{, \emph{i.e.}, all elements are in the ranking at all times.}  If, on the other hand, competitors or
teams enter or leave the ranking with time, $\Gamma$ grows and $\Omega$
decreases.  Therefore, the closure index indicates how \change{closed the system
is}. In Fig. \ref{fig:normalized} the values of $\Omega$ for the twelve
sports are given. The sports with the smallest closure are NASCAR-B, ESE, FCWR-G, GPI, and NASCAR-W. This means that in these datasets, there is a high rotation of elements for each time interval: it is common to have players or teams enter and leave the rankings. 
Datasets with a low $\Omega$ tend to have a high rank diversity, sometimes even for low ranks. 
The highest closure is measured for FIFA, FIDE-M, and FIDE-F. In these cases, teams and players tend to stay in the rankings over time.
Datasets with a high closure (close or equal to one) tend to have rank diversity curves that decrease towards the end of the ranking, as seen for FIFA, so the sigmoid $\Phi$ does not fit well (see below).

As discussed in reference~\cite{cocho2015rank}, the rank diversity can be
approximated by
\begin{equation}
\Phi_{\mu,\sigma}(\log k)=\frac{1}{\sigma\sqrt{2\pi}}
   \int_{-\infty}^{\log k}\exp\left(-\frac{(y-\mu)^2}{2\sigma^2}\right) {\rm d} y,
   \label{eqn:Phi}
\end{equation}
that is the cumulative of a Gaussian distribution. As will be seen in Fig.
\ref{fig:normalized} (left), $\Phi$ is close to the rank diversity for all sports
except ESE; this might be due to the poor statistics in this case. We can also
see that the model fails for FIFA, since $\Phi$ is monotonous while the rank
diversity falls for large $k$; the low value of $R^2$ ratifies this fact.

It is interesting to note that datasets with a poor $R^2$ value have either high or low values of $\Omega$, although there are examples of high or low $\Omega$ with a good fit, such as FCWR-G.

If the change of variable $k$ to $(\log(k)-\mu)/\sigma$ is performed, $\mu=0$
and $\sigma=1$ and what is called the unitary normal cumulative distribution
function is obtained. On the other hand, the empirical data also follow this
function, as shown in Fig. \ref{fig:normalized} (right).  A generic behavior has been
obtained. Since the rank diversity of all studied sports and games can be fitted with $\Phi$, it could be concluded that the rank dynamics have similar behaviors, independently on the nature and competitiveness of the sport and its ranking method.

We now consider three other statistical measures to study rank
dynamics~\cite{Ngramas}: the
change probability $p(k)$, the rank entropy $E(k)$ and the rank complexity
$C(k)$. The first is the probability that an element with rank $k$ changes its
rank. If for each sport we have ranking information for $T$ times
$t_1,t_2,\cdots,t_T$,
\begin{equation}
p(k)=\frac{\sum_{t=0}^{t=T-1} 1-\delta(X(k,t),X(k,t+1))}{T-1},
\label{eqn:rankChange}
\end{equation}
where $\delta$ denotes the Kronecker delta. 
Thus, if $\delta(X(k,t),X(k,t+1))=0$ a change in the occupation of rank $k$ has
occurred. For example, if two soccer teams alternate constantly between first
and second place every week, then the change probability will be highest
$p(1)=1$, as there is a rank change every time interval. However, for the same
hypothetical case, the rank diversity would be low $d(1)=d(2)=2/T$.

\begin{figure} 
\centering
\includegraphics[width=\textwidth]{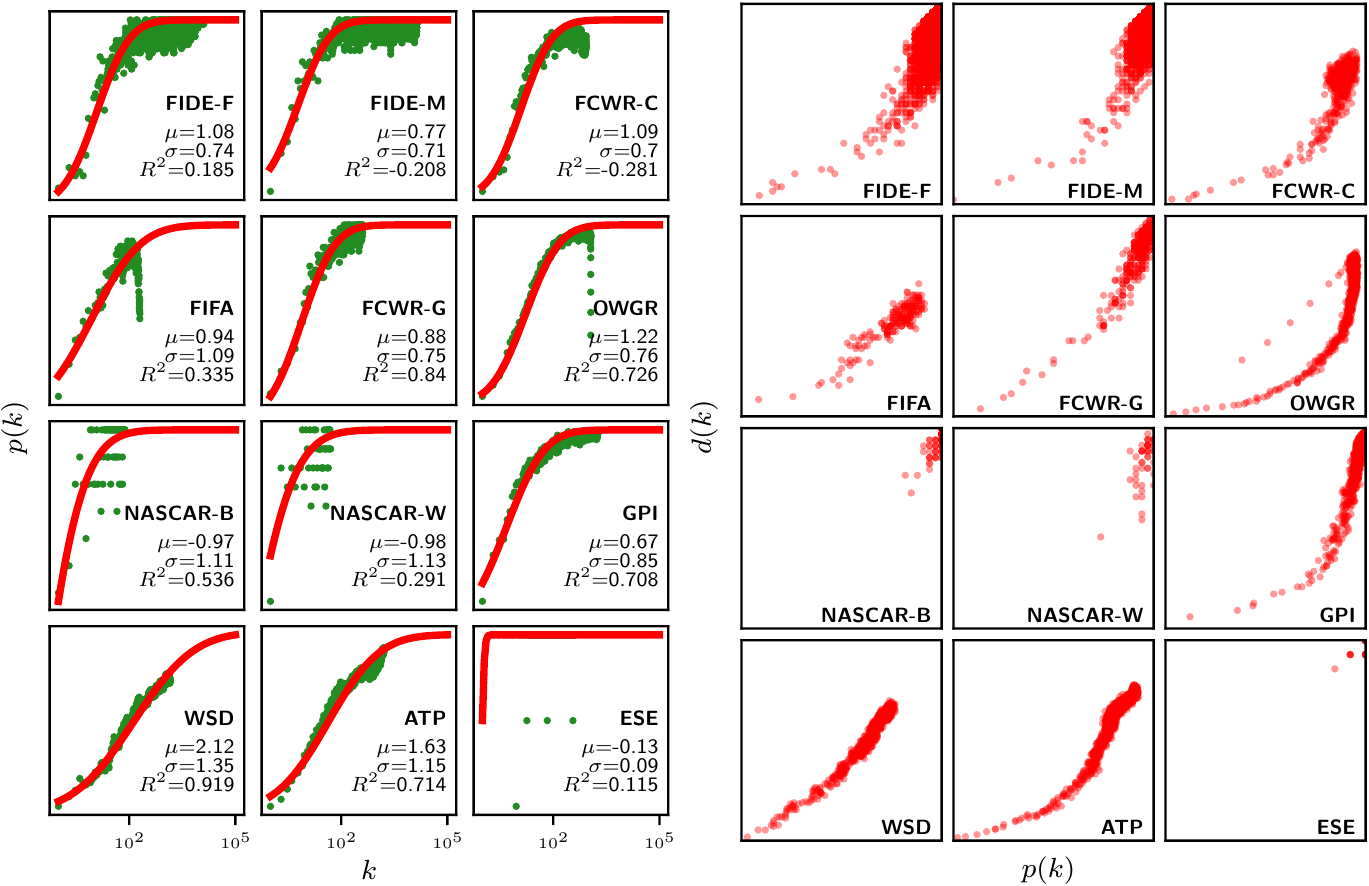}
\caption{(left) Change probability $p(k)$ (green dots) for the twelve rankings, together with a fitted sigmoid $\Phi$ (equation~\ref{eqn:Phi}) (red curve). The mean $\mu$, standard deviation $\sigma$, and squared error $R^2$ of $\Phi$ are also shown. (right) $p(k)$ vs. $d(k)$ plots.
The values of $d(k)$ and $p(k)$ are always in the interval $[0,1]$.}
\label{fig:RDvsRC}
\end{figure}

The values of $p(k)$ for the twelve rankings are given in Fig.~\ref{fig:RDvsRC}
(left). 
Within statistical error, all datasets follow closely the sigmoid $\Phi_{\mu,\sigma}(k)$ (equation~\ref{eqn:Phi}) except, in some cases, at the very end. 
The
values of $\mu$ and $\sigma$ for each dataset were obtained by fitting the
sigmoid to the data.

Fig.~\ref{fig:RDvsRC} (right) shows comparisons between change probability and
rank diversity. These show that the two measures are highly correlated,
although $p(k)$ increases consistently faster than $d(k)$, leading to convex
curves. For ESE and both NASCAR datasets, this is less clear because of few
data points.

We now define the rank entropy, $E(k)$. Let $k$ be an arbitrary rank and
let us consider an element $X$ contained in $X(k)$. The probability $p_X$ that $X$
occupies rank $k$ is equal to the number of times that $X$ had rank $k$ divided
by $T$. The rank entropy $E(k)$ is defined by
\begin{equation}
E(k)=-K\sum_{X\in X(k)}p_X \log p_X,
\label{eqn:entroprank}
\end{equation}
where $K$ is a normalizing constant given by $K=1/\log |X(k)|$,
so that $E(k) \in [0,1]$.  The rank entropy $E(k)$ quantifies the disorder of
the information we have with respect to the rank occupation. In other words, it
indicates how much information (disorder) will be obtained with new data. If
$E(k)=1$, it means that all the information about new ranks will be
``received'' with new data. If $E(k)=0$, it means that with previous
information we can predict the future, so new data does not carry any new
information~\cite{Shannon1948}.

\begin{figure} 
\centering
\includegraphics[width=\textwidth]{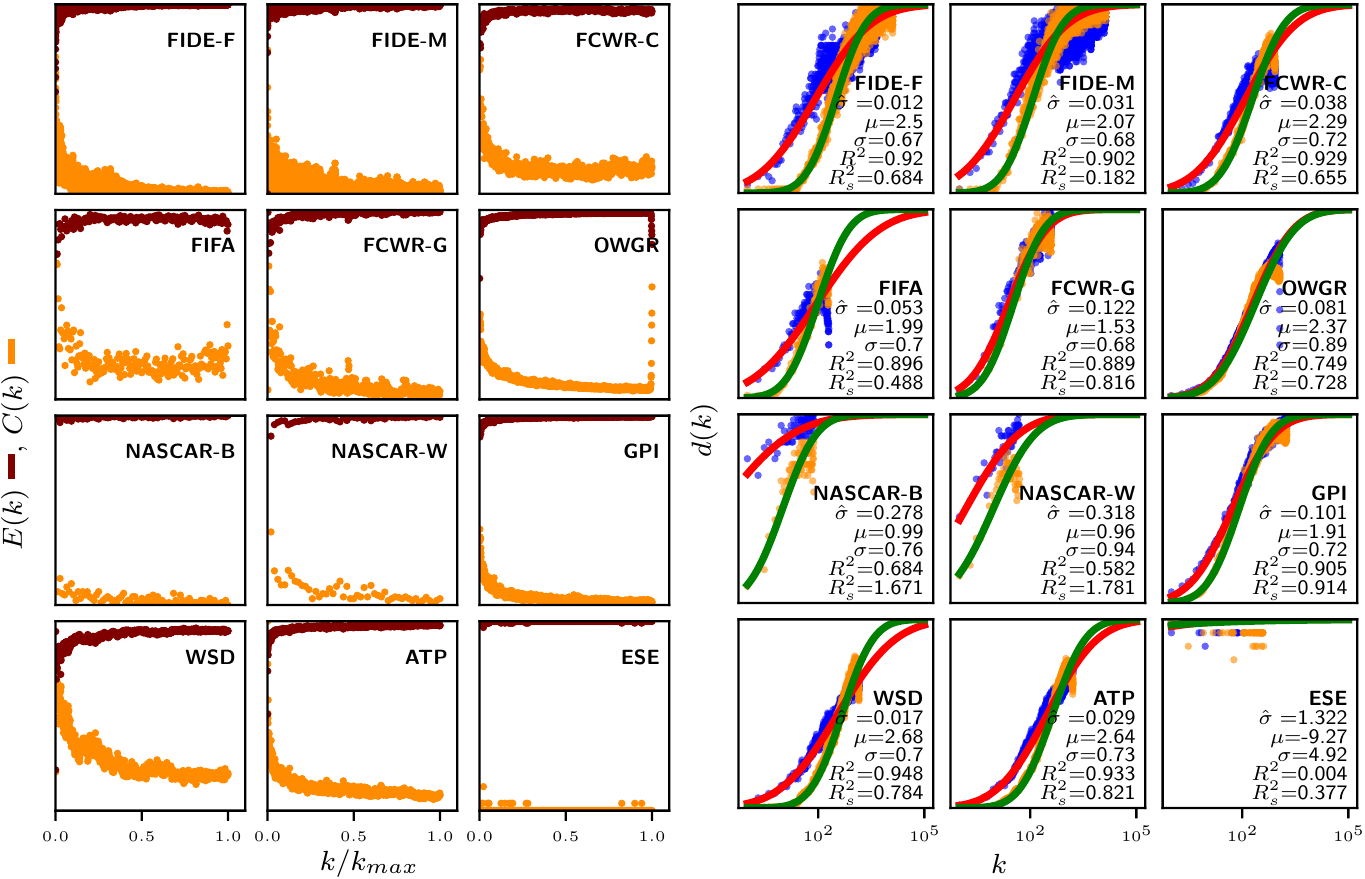}
\caption{(left) Rank entropy (red) and rank complexity (orange). Entropy is high, except for low ranks in some sports. Thus, complexity is low, except for low ranks in the same sports.
(right) Rank diversity $d(k)$ for both the twelve sports (blue dots) and the random walk model (orange
dots) The sigmoid fit $\Phi$ (eq. \ref{eqn:Phi}) is shown in red for the data and in green for the model.  The $\hat{\sigma}$ is the one used by the model (eq. \ref{eqn:randomWalk}) and is fine-tuned in such a way that the rank diversity of the
synthetic data approaches the best to the one of the empiric data. The $\mu$ and $\sigma$ refer to the $\Phi$ of the model (greeen sigmoid), while the $R^2$ is the difference between the model (orange dots) and its fit (green sigmoid). $R^2_s$ measures the difference between the data (blue dots) and the model (orange dots). 
The values of $d(k)$ , $E(k)$ and $C(k)$ are always in the interval $[0,1]$.}
\label{fig:entropy}
\end{figure}

In Fig. \ref{fig:entropy} (left) the rank entropy for the twelve systems is
given. We observe that in most games rank entropy is close to 1, which
indicates that in these systems, predictability is low and the rank occupation
is not stable, \emph{i.e.}, it is chaotic. It is to be noted that, in contrast to $d(k)$ and
$p(k)$, rank entropy shows that intermediate ranks have also a large amount of
disorder. Still, lower ranks are slightly more predictable. In other words, it
is easier (but still very limited) to predict who will be the first in a
ranking (usually the safest bet would be that there would be no change) than to
predict who would be in rank 100. 

Following~\cite{LopezRuiz:1995}, we can define rank complexity  as~\cite{fernandez2014information}:
\begin{equation}
C(k)=4\cdot E(k)\cdot (1-E(k)).
\label{eqn:complexrank}
\end{equation}
This represents a balance between ``order'' ($E(k)=0$) and ``disorder'' ($E(k)=1$),
since $C(k)$ is maximal (one) when $E(k)=0.5$ and minimal (zero) when the entropy is at
one of its extremes (zero or one). It has been argued that ``critical'' phenomena occur near a phase transition between order and chaos, and this coincides precisely with $C(k)=1$ \cite{GershensonFernandez:2012}. It can be said that $C(k)$ measures a ``balance'' between order and chaos.

 
In Fig. \ref{fig:entropy} (left) the graphs for $C(k)$, show that the rank
complexity is maximum when $E(k)\approx 0.5$. 
 It is clear that for FIDE, OWGR, FIFA, FCWR-G,
WSD and ATP, $C(k)$ is large for low $k$, \change{and} has a relative stability, which is
related to the small values of $d(k)$ and $p(k)$ for these ranks. In the other
systems, complexity is small
In particular, it will be seen that ESE shows for
all ranks the minimum values of $C(k)$ and the maximum values of $E(k)$. This,
together with what was obtained for $d(k)$ and $p(k)$, indicates that system
ESE is chaotic at the scale observed, there is no way to characterize it. It might also be because of lack of data, and more frequent rankings would yield more meaningful results.

\section{Random walk model} 

To reproduce the generic behavior of complex systems, the random walk model was
introduced in Ref.~\cite{morales2016generic}. In~\cite{cocho2015rank} it was
shown that the hierarchical structure of systems could be related to the presence
of random walks built with stochastic steps that follow a normal distribution.
Consider the map
\begin{equation}
\tilde k_{t+1} =  k_t + G(0, k_t \hat{\sigma}),
\label{eqn:randomWalk}
\end{equation}
where $G(0, \varsigma)$ represents the random number following Gaussian
distributions with null average and standard deviation $\varsigma$. 
Notice that the standard deviation in our model is proportional to the 
rank. This implies that elements with low
ranks vary less in time than the elements with large ranks. 
$\tilde k_{t+1}$ must be then unfold so that at time $t+1$ we obtain all
integers $1,2,\cdots,N$, albeit reordered. See~\cite{morales2016generic} for
more details. 

The results are
shown in Fig. \ref{fig:entropy} (right), where rank diversity for the
twelve sports is compared with the one obtained using the random walk model.
The blue points correspond to the data and the orange ones to the random model.
The red curve is the sigmoid $\Phi$ fitting the real data, while the green
curve is the sigmoid that approximates the model data, recall \eref{eqn:Phi}.
One can see that the curves
follow the data except for FIFA. This might be due to the fact that FIFA is the
most ``closed'' system, \emph{i.e.}, few new elements enter or exit the ranking throughout time. It is to be noted that for the lowest ranks, the real data
$d(k)$ is always larger than the model one.

\section{Discussion} 
By analyzing the rank dynamics of different sports, we found a common pattern:
the top ranked players or teams tend to change their rank slower than the rest.
This universal feature is independent of the nature of the sport, scoring
systems, and ranking criteria. It is clear that rank changes in each sport will
have different explanations. Still, the probability of change follows a common
pattern. Exploring the rank diversity, closure, change probability, rank
entropy, and rank complexity for different sports, we can find similarities
among them and also measure how ``fast'' ranks change depending on the sport.
This does not give us a strong predictive power, but allows us to infer how
fast changes occur in different sports, and also how predictable outcomes in
each competition are. 

It is clear that sports with higher values of rank diversity and rank entropy
imply that the ranks change faster. Prediction is limited, because it is not so
probable that the previous top players and teams will remain at the top. This
could be because chance is relatively more determinant than talent in the
outcomes in those sports, but also could be because those sports are more
competitive. Further work is needed to distinguish whether the shape of the
rank diversity curves can be used as a proxy for the ``randomness'' or
``competitiveness'' of each sport~\cite{Ben-Naim2006,Gabel2012}. 
In this direction, it would also be interesting to analyze how ``talent'' and ``luck'' contribute to success in different sports~\cite{Yucesoy2016,doi:10.1142/S0219525918500145,Pluchino2019,Janosov2020}. The ``Science of Success'' \cite{barabasi2019science} is still at an early stage, and it could benefit and be benefited from statistical studies of performance and success in sports.

\section*{Acknowledgments}
This work was supported by CONACyT under Grant  CB-285754;  and UNAM-PAPIIT under
Grants IG101421,  IN107919, and IV100120. 

%
%

\section*{References}
\bibliographystyle{ws-acs}
\bibliography{referencias}    
\end{document}